\documentclass{iopart}

\usepackage{epsfig}
\usepackage{here}
\usepackage{cite}

\begin{document}

\title[]{Experimental results on strangeness production 
         in proton-proton collisions at COSY}

\author{P.~Moskal$^{1,2}$\footnote{e-mail address: p.moskal@fz-juelich.de},
        H.-H.~Adam$^3$, A.~Budzanowski$^4$, R.~Czy{\.z}ykiewicz$^2$, D.~Grzonka$^1$,
        C.~Kolf$^1$, L.~Jarczyk$^2$,
        A.~Khoukaz$^3$, K.~Kilian$^1$, P.~Kowina$^{1,5}$, N.~Lang$^3$, T.~Lister$^3$,
        W.~Oelert$^1$, 
        C.~Quentmeier$^3$, R.~Santo$^3$, G.~Schepers$^{1}$, T.~Sefzick$^1$,
        M.~Siemaszko$^5$, 
        J.~Smyrski$^2$, A.~Strza{\l}kowski$^2$, 
        P.~Winter$^1$, M.~Wolke$^1$, P.~W{\"u}stner$^1$,
        W.~Zipper$^5$ 
      }

\address{ $^1$ IKP and ZEL, Forschungszentrum J\"{u}lich, D-52425 J\"{u}lich, Germany}
\address{ $^2$ Institute of Physics, Jagellonian University, PL-30-059 Cracow, Poland}
\address{ $^3$ IKP, Westf\"{a}lische Wilhelms--Universit\"{a}t, D-48149 M\"unster,Germany}
\address{ $^4$ Institute of Nuclear Physics, PL-31-342 Cracow, Poland} 
\address{ $^5$ Institute of Physics, University of Silesia, PL-40-007 Katowice, Poland}

\begin{abstract}
  The production of $K^+$ and $K^-$
  mesons in elementary proton-proton collision has been
  investigated at the Cooler Synchrotron COSY in J{\"u}lich.
  A high quality proton beam  with low emittance
  and small momentum spread permitted to study
  the creation of
  these mesons very close to the kinematical threshold.
 
  The energy dependence of the total cross section
  is investigated using internal beam facilities
  providing a high accuracy particle  momentum determination
  as well as an external non-magnetic  detection setup
  with a large geometrical acceptance.
 
  The determination of the four-momentum vectors
  for all ejectiles of each registered event
  gives the complete kinematical information allowing
  to study the interaction of the outgoing particles.
  Results on the performed studies of the
  $p p \to p p K^+ K^-$,
  $p p \to p \Lambda K^+$
  and $p p \to p \Sigma^0 K^+$ reactions
  will be presented and their relevance to the
  interpretation of heavy ion collisions will
  be discussed.
\end{abstract}




\section{Introduction}
\begin{figure}[H]
        \parbox{0.4\textwidth}{\epsfig{file=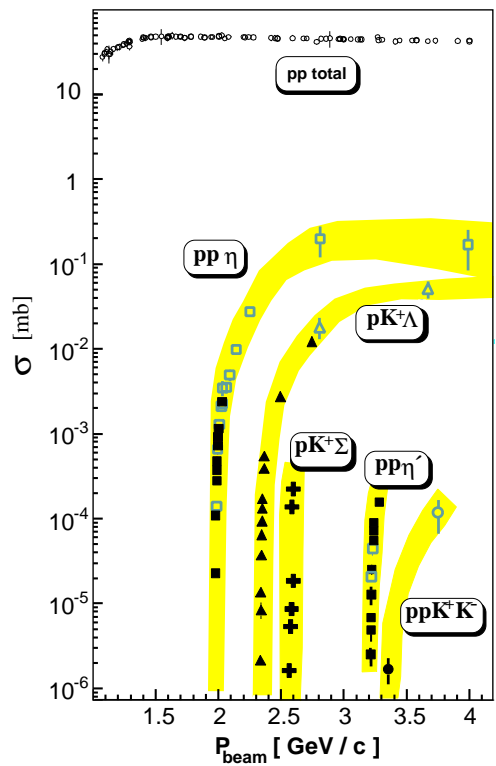,width=0.5\textwidth,angle=0}}
        \vspace{-0.4cm}
        \parbox{0.6\textwidth}{
        {\caption{\label{alltotal}\small  Close to threshold cross sections
            for the proton-proton interaction leading to the 
            production of mesons which wave function comprises significant
            amount of strangeness. For the comparison also the total reaction cross
            section of proton-proton collisions is shown.
            The filled points
            depict data taken at COSY~\cite{eta_smyrski,eta_menu,pkl_balewski,
            pkl_sewerin,pkl_bilger,etap1,etap2,kk_quentmeier}, 
            and  open symbols
            show results from other laboratories~\cite{eta_calen,eta_chiavassa,
            pkl_fickinger,pkl_bierman,pkl_louttit,kk_balestra,pp}.
        }}}
\end{figure}

\begin{figure}[H]
        \vspace{-0.7cm}
        \parbox{0.30\textwidth}{\epsfig{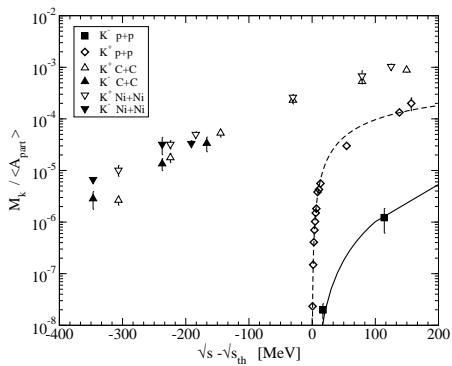}}
         \vspace{-0.1cm}
         \parbox{0.70\textwidth}{
         {\caption{\label{multi_kk}\small Multiplicity of kaon and anti-kaon
         production per participating nucleon
         for C+C~\cite{kaosCC}, Ni+Ni~\cite{kaosNiNi_1,kaosNiNi_2,Senger1}, and 
         proton-proton collisions
         \cite{pkl_balewski,pkl_sewerin,pkl_bilger,pkl_fickinger,kk_balestra,kk_quentmeier}. 
         The particle multiplicity M$_k$
         is defined as $\sigma_k/\sigma_R$, where $\sigma_R$ is the geometrical
         cross section equal to $\sigma_R$~=~0.94~b~\cite{kaosCC} 
         and $\sigma_R$~=~2.9~b~\cite{kaosNiNi_2}
         for carbon and nickel collisions, respectively.
         In case of
         proton-proton collisions $\sigma_R$ was taken as a total production
         cross section  equal to 45~mb. Diamonds represent the data for the
         $pp \to pK^{+} \Lambda$ reaction with the dashed line showing calculations
         assuming a constant primary production amplitude
         with the energy dependence defined by the phase space, proton-K$^+$
         Coulomb repulsion and a proton-$\Lambda$ strong interaction
         taken into account according 
         to reference~\cite{pkl_epj_balewski}.
         The solid line is described in the text.
        }}}
\end{figure}

\begin{figure}[H]
  \parbox{0.3\textwidth}{\epsfig{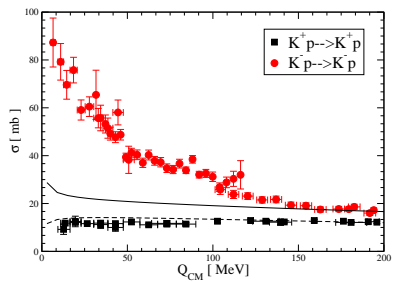}}
  \vspace{-0.4cm}
  \parbox{0.65\textwidth}{
  {\caption{\label{kpkp}{\small Cross section for the 
         K$^+$-proton\cite{goldhaber62}
         and K$^-$-proton\cite{adams75}
         elastic scattering.
         Solid and dashed lines represent the changes of phase space integral
         modified by the initial and final state Coulomb interaction only.
         Both curves were normalized to  points of large excess energies.
  }}}}
\end{figure}

One of the main topics of this conference is the study of the properties
of nuclear matter at high densities
realized via  relativistic
heavy ion collisions. The experimental studies are performed
by observation of 
the abundance and the phase space distributions of the produced
particles~\cite{senger209}. Information
about processes occuring during such collisions
are gained, in particular, by
the registration of the
$K^{+}$ and $K^{-}$ mesons~\cite{kaosCC,kaosNiNi_1,kaosNiNi_2},
which were created in the fireball region.
However, in order to learn 
-- from the observed kaon yields and momentum distributions --
about dense baryonic matter
and  properties
of strange particles immersed in it, 
the knowledge of their creation in the
elementary nucleon-nucleon collisons is indispensable.
To the same extent 
the information of their interaction with hadrons is important,
since being created in the dense nuclear environment they are likely
to undergo further reactions before being registered in detection systems.
The purpose of this talk is to give a short overview
of experimental achievements in the field of the K$^{+}$
and K$^{-}$ meson production in elementary proton-proton collisions
close to the corresponding kinematical threshold, where due to the 
rapid growth of the phase space volume available to the produced particles
the total cross sections increase by orders of magnitudes
over a few MeV range of excess energy as depicted in Figure~\ref{alltotal}.
Such studies have been
made possible due to the low emittance  and  small momentum spread proton beams
available at the storage ring facilities and in particular at the cooler 
synchrotron COSY.
The primary motivation for the investigation of  
hyperon and kaon production  in  elementary proton-proton
collisions close to threshold -- where only one partial wave dominates 
the reaction -- is i) the understanding of the creation mechanism
in the energy regime where both the hadronic and constituent quark-gluon
degrees of freedom may be relevant~\cite{kleefeld_gg}, and ii) the study of low energy 
hadronic interactions between nucleons and strange mesons or hyperons.
However, also investigations in the field of heavy ion collisions
could benefit from the results of the elementary processes.
A good example of the gain from  comparative studies of strangeness
production in heavy ion and elementary nucleon-nucleon collisions
is the observation that in nuclear matter kaons undergo a repulsion,
whereas
 anti-kaons feel a strong attractive potential. This leads to the
 splitting of their  effective masses~\cite{schaffner}.
The basis for this interpretation was 
the observation of the KAOS collaboration
that the multiplicity  of  kaons produced in the C+C or Ni+Ni collisions
is almost the same as for  anti-kaons at the corresponding
center-of-mass
energy above threshold with respect to  nucleon-nucleon kinematics.
 This was in a drastic contrast to   elementary
proton-proton collisions where the multiplicity of K$^+$ mesons
is determined to be two orders of magnitude
larger than  for anti-kaons,
as can be seen in Figure~\ref{multi_kk}, presenting additionally to the results of 
KAOS  data taken at  
the laboratories COSY, SATURNE and BNL.
 The consequences of this comparison become even more astonishing
 if one takes into account that in nuclear matter kaons
 are much  less absorbed than anti-kaons, since the latter 
 in  collisions with  nucleons form  hyperon resonances
 which may subsequently decay weakly into a pion-nucleon system.
 It is worth to note, that the  repulsive hadronic interaction
 of kaons with protons and the attractive interaction
 of anti-kaons and protons
 is also seen in the shape of the  energy dependence of the cross section
 for K$^+$p or K$^-$p elastic scattering presented in Figure~\ref{kpkp}.
 When comparing the data
 to calculations -- including the changes of phase-space integral and 
 Coulomb interaction in
 the initial and final states --
 one observes a huge enhancement for
 the K$^-$p cross section with decreasing excess energy
 and a slight suppression in case of  K$^+$p scattering.
 This observation may be attributed to the slight repulsion
 due to the kaon-proton hadronic interaction and significantly
 larger attraction caused by the strong
 interaction between the K$^-$  and proton due to the vicinity of the
 $\Lambda(1405)$ hyperon resonance. The effect  must
 have a direct influence on the kaon mass splitting 
 in dense nuclear matter.
 A natural step on the way to understand the medium modification
 of strange meson properties
 is to study how the presence of a second
 nucleon or hyperon would influence the kaon-proton
 or anti-kaon proton interaction - an issue interesting in itself as well. 
 This can be studied by measuring
 the energy dependence of the total cross section
 for the  reactions $pp\to pK^{+}\Lambda$ 
 or $pp\to  ppK^{+}K^{-}$ close to their
 corresponding kinematical thresholds~\cite{magnus_sibi}
 or via a study of
 the  distribution of  double differential cross sections,
 for example in the Dalitz-plot representation~\cite{dalitz} 
 which gives the complete, experimentally obtainable, information
 about reactions with three particles in a  final state~\cite{kilianproc}.
\vspace{0.2cm}

\section{Measurements of the $pp\to ppK^{+}K^{-}$ reaction near threshold at COSY}
 A primordial motivation for studying the $pp\to ppK^{+}K^{-}$ reaction
close to the kinematical threshold was described extensively by W. Oelert
at the first Cracow Workshop on Meson Production and Interaction~\cite{waltermeson}. 
It concerns the study of the hadronic interaction between K$^+$ and
K$^{-}$ mesons and in particular the investigations of the still 
unknown origin of the scalar resonances f$_0$(980) and a$_0$(980).
As a possible interpretation of their structure one considers
exotic four quark~\cite{f0_1},
conventinal $q\bar{q}$~\cite{f02_3},
or molecular like $K\bar{K}$ bound~\cite{f04_5} states. 
\vspace{0.0cm}
\begin{figure}[H]
        \parbox{0.35\textwidth}{\epsfig{file=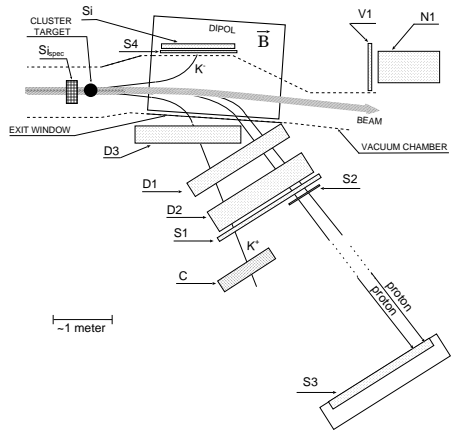,width=0.45\textwidth,angle=0}}
        \parbox{0.6\textwidth}{
         {\caption{\label{detector}\small  Schematic view of the COSY-11 
           detection setup\cite{c11_brauksiepe}. Only detectors needed for the
           measurements of the reactions 
           $pp\to ppK^{+}K^{-}$, $pp\to pK^{+}\Lambda(\Sigma^{0})$
           $pp \to nK^{+}\Sigma^{+}$ and $pn \to n K^{+} \Lambda(\Sigma^{0})$ are shown. 
           D1, D2, D3, denote the drift chambers, S1, S2, S3, S4 
           and V1 the scintillation detectors;
           N1 the neutron detector, C $\check{\mbox{C}}$erenkov counter, 
           and Si and Si$_{spec}$ silicon strip detectors.
        }}}
\end{figure}
The first experiments aiming at the determination of the total cross section
for the $pp\to ppK^{+}K^{-}$ reaction very close to threshold
showed however that it is more than seven orders
of magnitude smaller than the total proton-proton production 
cross section (see fig.~\ref{alltotal})
and hence the extraction of the signals required  thorough investigations
of possible background reactions~\cite{magnus_stori,
                                       phd_magnus,phd_lister,
                                       phd_quentmeier}.
 The experiments are being performed at the cooler synchrotron COSY~\cite{cosy_dieter},
using the COSY-11 detection system shown in Figure~\ref{detector}
and a hydrogen cluster target~\cite{dombrowski}
installed in front of one of the regular COSY dipole magnets as depicted schematically
in Figure~\ref{detector}. The target being a beam of H$_2$ molecules grouped to clusters
of up to $10^{6}$ atoms crosses perpendicularly the COSY beam 
with intensities up to~$5\cdot 10^{10}$
protons. The very thin cluster target of only $10^{14}$ atoms/cm$^{2}$
makes the probability of secondary scattering negligible and hence allows
the precise determination of the ejectile momenta.
However, despite the low density of the target it is still possible to measure reactions 
whith cross sections in the nanobarn region,
since the proton beam circulating in the ring hits the target more than
$10^{6}$ times per second 
resulting in  luminosities of up to~$5\cdot 10^{30}$~cm$^{-2}$s$^{-1}$.
If at the intersection point of the cluster target and COSY beam 
a collision of protons leads to the production of a K$^{+}$K$^{-}$ meson pair,
then the ejected particles -- having smaller momenta than the circulating beam --
are directed by the magnetic field 
towards the detection system and leave the vacuum chamber through
the thin exit foils~\cite{c11_brauksiepe}. Tracks of the positively charged particles,
registered by the drift chambers, 
are traced back through the magnetic field
to the nominal interaction point, leading to a momentum determination. 
A simultanous measurement of the velocity, performed by means of 
scintillation detectors, 
permits to identify the 
registered particle and to determine  its four momentum vector.
Since at threshold the center-of-mass momenta of the produced particles
are small in comparison to the beam momentum,  in the laboratory
all ejectiles are moving with almost the same velocity. This means that the 
laboratory proton momenta are almost two times larger then the momenta of 
kaons and therefore in the dipole magnetic field protons expierience a much larger 
Lorentz force than kaons.
As a consequence, in case of the near threshold production,
protons and kaons are registered in 
separate parts of the drift chambers as shown schematically in Figure~\ref{detector}.
 Figure~\ref{invmass}(left) shows the squared mass of two simultaneously detected
particles in the right half of the drift chamber.
A clear separation is observed into groups of events with two protons,
two pions, proton and pion and also deuteron and pion.
This spectrum enables to select events with two registered protons.
The additional requirement that the mass of the third particle,
registered at the left side of the chamber,
corresponds to the mass of the kaon 
allows to identify  events with  a $pp\to ppK^{+}X^{-}$ reaction signature.
Knowing both the four momenta of positively charged ejectiles
and the proton beam momentum
one can calculate the mass of an unobserved system $X^{-}$. 
Figure~\ref{invmass}(middle) presents the square of the missing mass spectrum 
with respect to the identified ppK$^{+}$ subsystem.
In case of the $pp\to pp K^{+}K^{-}$ reaction this should
correspond to the mass of the $K^{-}$ meson, and indeed 
 a pronounced signal can be clearly recognised.
\begin{figure}[H]
       \vspace{-0.6cm}
       \parbox{0.35\textwidth}{\epsfig{file=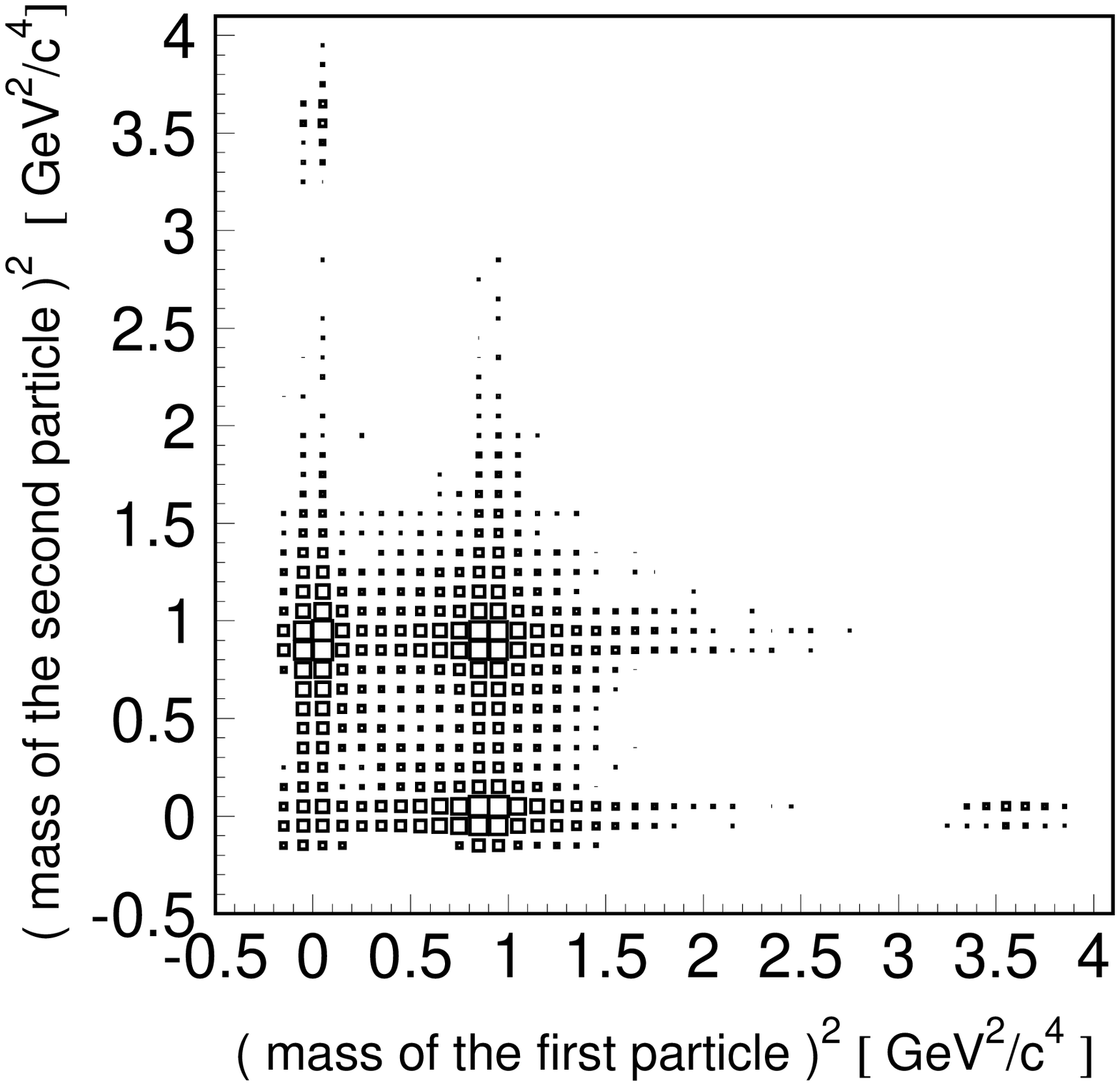,width=0.4\textwidth,angle=0}}
       \parbox{0.3\textwidth}{\vspace{0.2cm}\epsfig{file=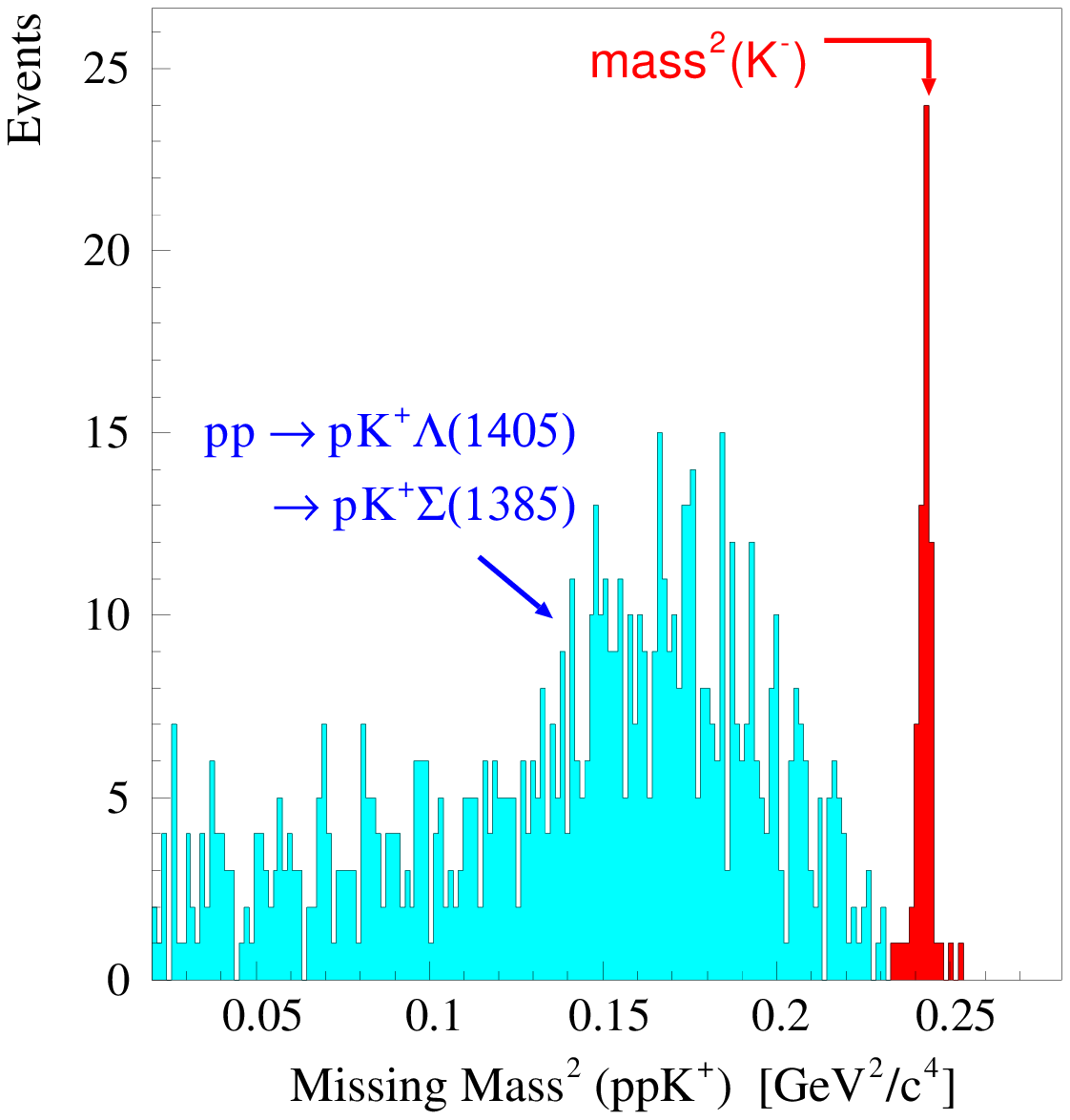,width=0.3\textwidth,angle=0}}
       \parbox{0.25\textwidth}{\epsfig{file=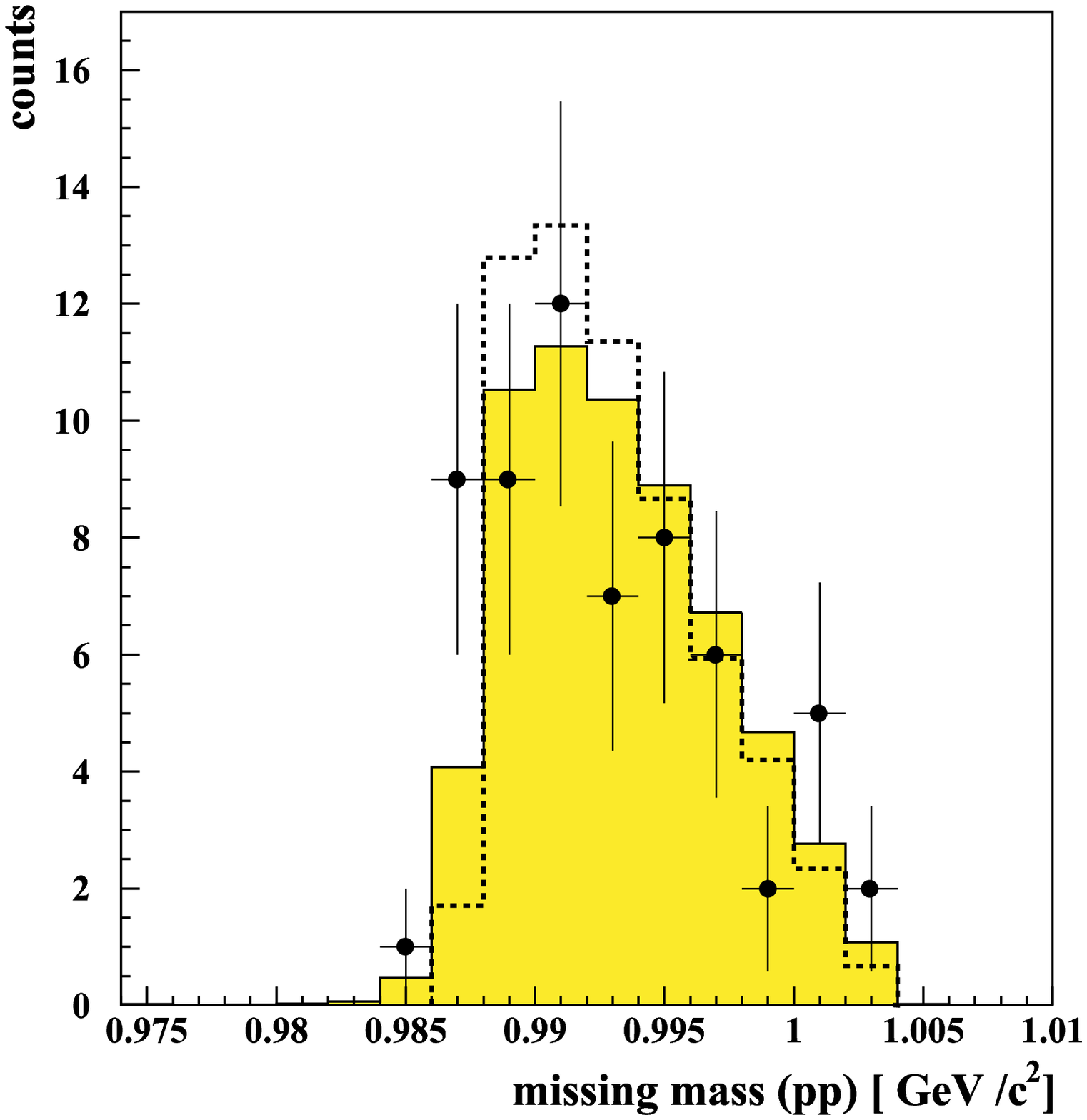,width=0.37\textwidth,angle=0}}
       \vspace{-0.4cm}
        {\caption{\label{invmass}\small 
          (left) Squared masses of two positively charged particles measured in coincidence
            at the right half of the drift chambers~(see fig.\ref{detector}).
            The number of events is shown in logarithmic scale. \ \
          (middle) Missing mass spectrum determined for the
            $pp\to ppK^{+} X^{-}$ reaction at an excess energy of Q~=~17 MeV~\cite{kk_quentmeier}.
          (right)   Experimental  spectrum of the K$^{+}$K$^{-}$
            pair invariant mass measured for the
            $pp\to ppK^{+}K^{-}$ reaction. 
        }}
\end{figure}
The additional broad structure seen in the Figure is partly due to the
$pp\to pp\pi^{+}X^{-}$ reaction, where the $\pi^{+}$ was misidentified
as a K$^{+}$ meson but also
due to K$^{+}$ meson production associated with the hyperons 
$\Lambda$(1405) or $\Sigma$(1385), e.g. via the reaction
$pp\to pK^{+}\Lambda(1405)\to pK^{+}\Sigma\pi \to pK^{+}\Lambda\gamma\pi \to pK^{+}p\pi\gamma\pi$.
In this case the missing mass of the ppK$^{+}$ system corresponds to the
invariant mass of the $\pi\pi\gamma$ system and hence can acquire values
from twice the  pion mass up to the kinematical limit.
This background, however, can be completely reduced by demanding a signal 
in the silicon strip detectors at the position where the K$^{-}$
meson originating from the $pp\to ppK^{+}K^{-}$ reaction is expected~\cite{kk_quentmeier}.

This clear identification allows to determine the total cross section and the corresponding
multiplicity shown in Figures~\ref{alltotal} and \ref{multi_kk}, respectively. 
The solid line in Figure~\ref{multi_kk}
         represents results of calculations~\cite{sibi_cas} for the $pp\to ppK^{+}K^{-}$
         reaction taking into account the changes of the production amplitude 
         as deduced from the K$^{+}$p and K$^{-}$p elastic scattering shown 
         in Figure~\ref{kpkp},
         but neglecting 
         the influence of the dominant proton-proton interaction!
On the other hand the proton-proton FSI in case of three body final states
e.g.\ $pp\to pp\pi^{0}$~\cite{pi_meyer,pi_uppsala}, $pp\to pp\eta$~\cite{eta_calen,eta_smyrski},
or $pp\to pp\eta^{\prime}$~\cite{etap2,swave} influences strongly the total cross section 
energy dependence by enhancing it by more than an order of magnitude  for
excess energies  below  Q~$\approx$~15~MeV. 
Thus it is surprising that in spite of its neglection one can describe
the data of the $pp\to ppK^{+}K^{-}$ reaction. The origin of that effect 
will be investigated experimentally in the near future~\cite{magnusproposal}.
At present one can only speculate whether it is 
due to the partial compensation of the pp and K$^{-}$p hadronic
interaction or maybe by the additional degree of freedom
given by the four body final state~\cite{magnus_sibi}.
At present it is also not possible to judge to what extent the
close to threshold production of K$^{+}$K$^{-}$ pairs proceeds
through the intermediate doorway state f$_{0}$(980).
The experimentally determined distribution of the missing mass with respect to 
the proton-proton system is shown in Figure~\ref{invmass}(right) 
and demonstrates that the non-resonant K$^{+}$K$^{-}$
production (shaded area) is hardly distinguishable from the resonant
$pp\to ppf_{0}(980)\to ppK^{+}K^{-}$ reaction sequence 
(dashed line)~\cite{phd_quentmeier}. 
It is clear that the up to date statistics of the 
data is not sufficient to favour one of the two processes.
Recently, the mass spectrum of the K$^{+}$K$^{-}$ pair 
produced via the $pp\to ppK^{+}K^{-}$ reaction 
was calculated in the framework of the $\pi\pi - \mbox{K}\overline{\mbox{K}}$ 
model of the J{\"u}lich group for cases where f$_{0}$
is a genuine meson  or $\mbox{K}\overline{\mbox{K}}$
 bound state~\cite{haidenbauer_proc}.
However, again the present statistics is not enough to distinguish
between the two hypotheses.

\section{Study of the $pp\to pK^{+}\Lambda$ reaction at COSY}
\begin{figure}[H]
        \vspace{-0.4cm}
        \begin{center}
        \hspace{2cm}
        \parbox{0.75\textwidth}{\epsfig{file=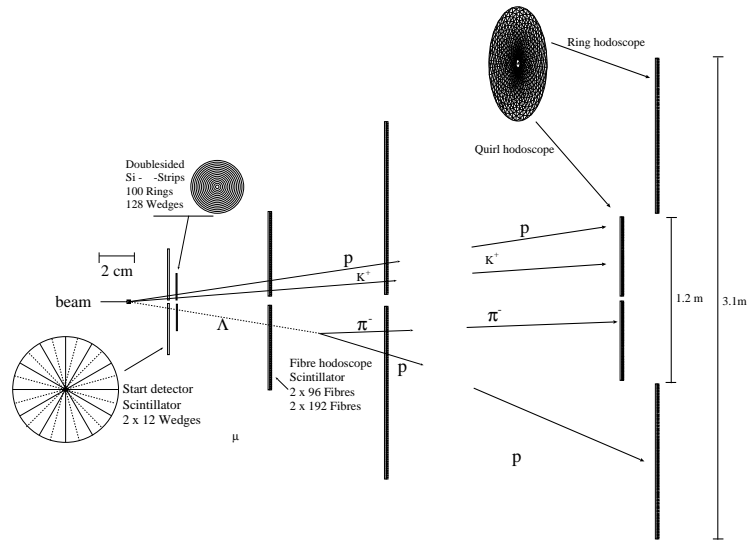,width=0.75\textwidth,angle=0}}
        \end{center}
         {\caption{\label{tofdetector}\small 
          Schematic view of the TOF detection system with  superimposed tracks 
          from the typical $pp\to pK^{+}\Lambda\to pK^{+} p \pi^{-}$ reaction sequence.
        }}
\end{figure}

 The kaon production associated with the hyperon $\Lambda$ or $\Sigma^{0}$
is studied experimentally very close to threshold 
at the internal facility COSY-11~\cite{pkl_old_balewski,pkl_balewski,pkl_sewerin}
presented in the 
previous section and also complementary for the higher energies at the
large acceptance non-magnetic time-of-flight spectrometer COSY-TOF~\cite{pkl_bilger,tof_strange}
which functioning shall be presented below.
The TOF detector consists of four rotationally symmetric 
detection layers positioned close to the 
LH$_{2}$ liquid hydrogen target ~\cite{tof_target} 
and the "Quirl" scintillation hodoscope~\cite{tof_dahmen} as shown schematically
in Figure~\ref{tofdetector}. To minimize  multiple scattering 
of beam protons and reaction products the whole system is enclosed in a vacuum vessel
consisting of 3~m outer diameter  barrel elements providing 
up to 8~m flight path in vacuum. 
An event from  a production of a neutral hyperon -- with a  mean decay path of a few cm --
gives  only two signals in the target start detectors and four in the  
fiber ring hodoscopes as is depicted in Figure~\ref{tofdetector}.
This allows to use  a very selective trigger based on the hit multiplicity in the
scintillation detectors. Hit positions measured in the start detector, 
in the double sided micro-strip silicon detector, the  four layers of fiber scintillators
and in the "Quirl" or "Ring" scintillation hodoscopes
permit the precise reconstruction of the primary and decay vertices.
From the reconstructed tracks after a mass assignment  four-momentum vectors
of all registered particles may be determined for each event.
Due to the possibility of monitoring the beam-target luminosity 
both COSY-11 and TOF experiments can establish
not only cross section distributions but also their absolute magnitude.

 Gernerally the knowlegde of four-momentum vectors of all ejectiles,
for each event,
allows to study the low energy interaction among the produced particles,
and specifically the hyperon-nucleon or kaon-nucleon interaction.
A quantitative study of the Dalitz-plot occupation 
in case of the $pp\to pK^{+}\Lambda$ reaction performed by the COSY-11 
collaboration resulted already in 
the estimation of scattering length and effective range parameters
averaged over the spin states, with 
values of 2~fm and 1~fm being extracted, 
respectively~\cite{pkl_epj_balewski}.

With the installation of neutron detectors~\cite{tof_neutron,c11_neutron}
both the
TOF and COSY-11 facility allow also to study charged hyperon production,
for example $\Sigma^{+}$ via the $pp \to n K^{+} \Sigma^{+}$ reaction.
In case of the COSY-11 setup the neutron detector is positioned as shown
in Figure~\ref{detector} and in case of the TOF facility it is installed 
 behind the "Quirl" scintillation detector.
Using the TOF facility the $pp\to n K^{+}\Sigma^{+}$ reaction can 
 again be identified by the event topology
combined with the mass hypothesis and the missing mass method,
 whereas in case of COSY-11
the measurements of kaons and neutrons will allow to   
recognise the reaction via the  missing mass technique only.

Data on the close to threshold kaon production are necessary
in order to learn about the strangeness production mechanisms.
Especially, they are needed to verify the hypothesis
of the destructive interference between $\pi$ and K meson exchanges~\cite{gasparian},
proposed to explain the COSY-11 observation that
close to threshold
the $K^{+}$ meson, when associated with the hyperon $\Lambda$ 
 is by a factor of $\sim$30 more copiously produced than  when
 created together with a $\Sigma^{0}$ hyperon~\cite{pkl_sewerin}. 

Presently, an additional tracking system consisting of 
straw chambers and silicon micro-strip detectors 
is being built for the TOF installation, which will improve
the tracking possibilities and as a consequence allow
for the registration of more complex
decay patterns originating e.g. 
from the production of hyperons belonging to the 3/2 decuplet~\cite{dieterkilian}.
For instance, the production of the 
$\Sigma^{*+}(1385)$ hyperon associated with the K$^{0}$
meson can be identified by the reconstruction of 
the $\Sigma^{*+}\to \pi^{+} \Lambda$ prompt decay geometry and the delayed
decays with separate vertices $\Lambda \to p\pi^{-}$ 
and $K^{0} \to \pi^{+}\pi^{-}$~\cite{dieterkilian}.
The upgrade of the COSY-11  detection system by a new hexagonal drift chamber~\cite{hex}
and a $\check{\mbox{C}}$erenkov detector~\cite{cerenkov} increases the efficiency of the  
kaon detection and its distinction from pions, which together with a 
near future installation 
of a  spectator detector~\cite{spec_proc} will allow to 
measure close to threshold cross sections for 
the quasi-free $pn\to n K^{+} \Lambda(\Sigma^{0})$ reactions.

\end{document}